\documentclass[conference]{IEEEtran}
\IEEEoverridecommandlockouts
\usepackage{cite}
\usepackage{amsmath,amssymb,amsfonts}
\usepackage{algorithmic}
\usepackage{graphicx}
\usepackage{textcomp}
\usepackage{xcolor}
\usepackage{array}

\def\BibTeX{{\rm B\kern-.05em{\sc i\kern-.025em b}\kern-.08em
    T\kern-.1667em\lower.7ex\hbox{E}\kern-.125emX}}

\begin{document}

\title{CAG-Avatar: Cross-Attention Guided Gaussian Avatars for High-Fidelity Head Reconstruction}%

\author{
\IEEEauthorblockN{
Zhe Chang\IEEEauthorrefmark{1},
Haodong Jin\IEEEauthorrefmark{1},
Yan Song\IEEEauthorrefmark{1},
Hui Yu\IEEEauthorrefmark{2}}
\IEEEauthorblockA{\IEEEauthorrefmark{1}Dept. of Control Science and Engineering, University of Shanghai for Science and Technology, Shanghai, China\\
Email: 233370890@st.usst.edu.cn, 231260086@st.usst.edu.cn, sonya@usst.edu.cn}
\IEEEauthorblockA{\IEEEauthorrefmark{2}School of Psychology and Neuroscience, University of Glasgow, Glasgow, UK\\
Email: hui.yu@glasgow.ac.uk}
}

\maketitle

\begin{abstract}
Creating high-fidelity, real-time drivable 3D head avatars is a core challenge in digital animation. While 3D Gaussian Splashing (3D-GS) offers unprecedented rendering speed and quality, current animation techniques often rely on a "one-size-fits-all" global tuning approach, where all Gaussian primitives are uniformly driven by a single expression code. This simplistic approach fails to unravel the distinct dynamics of different facial regions, such as deformable skin versus rigid teeth, leading to significant blurring and distortion artifacts. We introduce Conditionally-Adaptive Gaussian Avatars (CAG-Avatar), a framework that resolves this key limitation. At its core is a Conditionally Adaptive Fusion Module built on cross-attention. This mechanism empowers each 3D Gaussian to act as a query, adaptively extracting relevant driving signals from the global expression code based on its canonical position. This "tailor-made" conditioning strategy drastically enhances the modeling of fine-grained, localized dynamics. Our experiments confirm a significant improvement in reconstruction fidelity, particularly for challenging regions such as teeth, while preserving real-time rendering performance.
\end{abstract}


\section{Introduction}
Creating high-fidelity, drivable digital avatars from monocular video is a cornerstone for advanced interaction systems, ranging from VR/AR and telepresence to social vision in intelligent vehicles~\cite{yu2023social,guo2021adnerf,hong2022headnerf}. An ideal avatar must balance visual realism, ease of capture, and real-time performance. However, reconstructing and animating a human head remains a formidable challenge due to its complex geometry and dynamics. These challenges are amplified when considering the full spectrum of human expression, which includes not only typical emotions but also highly complex, asymmetrical movements found in conditions such as facial palsy~\cite{xia2022aflfp}. This highlights the need for models that can handle highly localized and non-uniform deformations.

Previous methods exhibit notable trade-offs. While 3D Morphable Models (3DMMs)~\cite{blanz1999morphable} capture global facial structure, their linear nature limits the modeling of fine details and complex deformations, particularly struggling to represent the subtlety and nuance of genuine emotional expressions~\cite{wang2023mgeed}. Neural Radiance Fields (NeRF)~\cite{mildenhall2021nerf} and its variants~\cite{gafni2021dynamic, guo2021adnerf, hong2022headnerf} achieve photorealism but suffer from prohibitive computational costs, making them unsuitable for real-time applications. Even accelerated NeRFs using explicit structures~\cite{xu2023avatarmav,zielonka2023instant} are still bottlenecked by the inherent inefficiency of volume rendering.

Recently, 3D Gaussian Splatting (3D-GS)~\cite{kerbl2023gaussian} revolutionized real-time, high-quality rendering by using an explicit, rasterization-based representation. This has spurred a new wave of research in creating animatable head avatars~\cite{shao2024splattingavatar,xiang2024flashavatar,xu2024gaussianhead,zhao2024psavatar}. However, some methods have a fundamental limitation in the animation mechanism, such as ~\cite{xiang2024flashavatar}. They employ a "one-size-fits-all" strategy by concatenating a global expression code with each Gaussian's features, forcing a single MLP to learn all region-specific responses. This approach has difficulty disentangling the mixed dynamics of the face, for example, the non-rigid deformation of the skin and the rigid motion of the teeth.

To overcome this bottleneck, we propose Conditionally-Adaptive Gaussian Avatars (CAG-Avatar), a novel framework that integrates the efficiency of 3D-GS with the power of the Cross-Attention Mechanism~\cite{vaswani2017attention}. At its core, we introduce a conditionally-adaptive fusion module. This module enables each Gaussian primitive to act as a \textit{Query}, adaptively attending to the global expression code (serving as \textit{Key} and \textit{Value}) based on its canonical position. This "tailor-made" conditioning for each primitive significantly improves the modeling of fine-grained local dynamics, especially for rigid components such as teeth.

Our main contributions are as follows:
\begin{enumerate}
    \item We propose a cross-attention-based fusion module that allows each Gaussian primitive to dynamically extract relevant driving signals based on its location, addressing detail loss and ambiguity from global conditioning.
    \item We design an efficient, high-fidelity head reconstruction framework that integrates this module with 3D-GS, significantly enhancing the modeling of mixed dynamics (e.g., skin vs. teeth) while maintaining real-time rendering performance.
\end{enumerate}

\section{Related Work}
\subsection{3D Head Avatar Reconstruction} 
The reconstruction of 3D facial avatars has evolved from traditional geometric to modern deep learning methods. A pioneering work is the 3DMM~\cite{blanz1999morphable}, which embeds facial shapes and appearances into a low-dimensional linear subspace. While many subsequent works~\cite{yu2012perception,guo20213d,zhang2023hack} have enhanced its expressive power, 3DMMs are fundamentally limited by their fixed mesh topology and linear representation, failing to capture high-frequency details like wrinkles or complex structures like hair. This has motivated the pursuit of more expressive, non-linear representations.

\subsection{NeRF-based Facial Avatars}
The advent of NeRF~\cite{mildenhall2021nerf} marked a paradigm shift by using an implicit neural function to synthesize photorealistic novel views. This powerful capability was quickly adopted for dynamic facial avatars~\cite{park2021nerfies,gafni2021dynamic,guo2021adnerf}. For instance, Gafni et al.~\cite{gafni2021dynamic} proposed a dynamic NeRF for monocular 4D reconstruction. However, NeRF's high computational cost, stemming from its dense ray sampling and volume rendering mechanism, makes it unsuitable for real-time applications. Even accelerated hybrid methods such as INSTA~\cite{zielonka2023instant}, which use explicit data structures, are still bottlenecked by the inherent inefficiency of ray marching. This has spurred the exploration of explicit representations that bypass volume rendering entirely.

\subsection{3D Gaussian Splatting for Dynamic Avatars}
3D-GS~\cite{kerbl2023gaussian} provides a revolutionary path to real-time, high-quality rendering utilizing an explicit representation of anisotropic 3D Gaussians and an efficient rasterization pipeline. This breakthrough was rapidly extended to dynamic scenes~\cite{yang2024deformable,wu20244d} and, more specifically, to drivable head avatars~\cite{xiang2024flashavatar,chen2024monogaussianavatar}. These methods typically animate the avatar by predicting the deformation of Gaussian primitives from a canonical space to a target expression, a process often guided by a parametric model such as FLAME.

However, as we highlighted in our introduction, current dynamic 3D-GS methods for head avatars, such as FlashAvatar~\cite{xiang2024flashavatar}, share a common theoretical bottleneck in their deformation modeling. These approaches typically employ a direct conditioning strategy where a global expression code is simply concatenated with each Gaussian's positional features. This spatially-invariant method applies a homogeneous driving signal to all primitives, ignoring the specific responses required by different facial regions. This simplified assumption is particularly inadequate for handling mixed dynamics---namely, the non-rigid deformation of skin versus the rigid motion of teeth.  It not only imposes a significant learning burden on the deformation network but also leads to visual artifacts like blurring and distortion in the oral region. Therefore, how to provide a more discriminative, position-adaptive driving signal for each Gaussian remains the core problem that our research aims to address.

\begin{figure*}[!ht]
\centering
\includegraphics[width=1\linewidth]{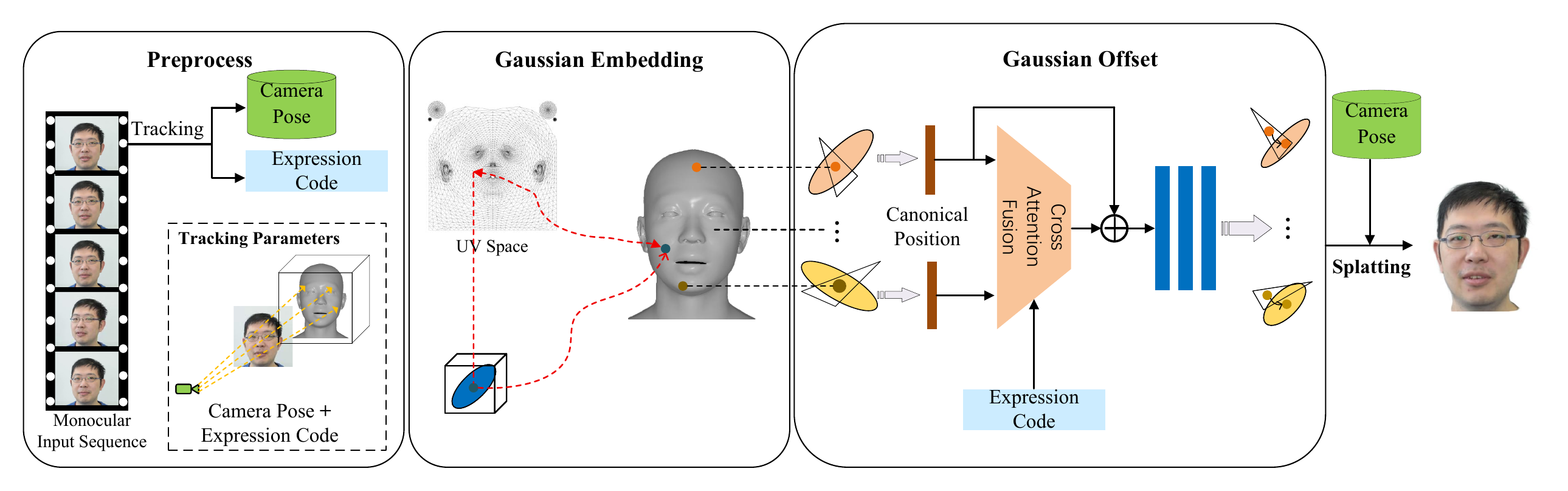}
\caption{Diagram of the proposed method. We represent the 3D head as a Gaussian field initialized in a 2D UV space and subsequently embedded onto a dynamic FLAME mesh surface via mesh rasterization. To animate the Gaussians, the canonical position of each Gaussian center is first encoded into a positional feature vector. This feature vector is then refined using the tracked expression code through a cross-attention fusion module, where positional features act as queries and the global expression code provides the key and value. The resulting expression-conditioned features are integrated with the original positional features via a residual connection. This fused representation is then fed into an offset MLP to predict the final spatial deformation, including offsets in position, rotation, and scale. Finally, the deformed Gaussians are splatted to render the portrait image under a given camera pose.}
\end{figure*}

\section{Method}
\subsection{Preliminary}
3D-GS~\cite{kerbl2023gaussian} presents a method for reconstructing static scenes from a set of images with known camera parameters. The scene is represented as a collection of anisotropic 3D Gaussians. Each Gaussian is defined by its center (mean) $\mu \in \mathbb{R}^3$ and a 3D covariance matrix $\Sigma$.

\begin{equation}
    G(\mathbf{x}) = e^{-\frac{1}{2}(\mathbf{x} - \boldsymbol{\mu})^{\top} \Sigma^{-1} (\mathbf{x} - \boldsymbol{\mu})}
    \label{eq:gaussian_pdf}
\end{equation}

Note, a covariance matrix $\Sigma$ is physically meaningful only if it is positive semi-definite, a property not guaranteed during optimization via gradient descent. To ensure differentiability and maintain this property, Kerbl et al.~\cite{kerbl2023gaussian} decompose the covariance matrix $\Sigma$ into a rotation matrix $R$ and a scaling matrix $S$. These are parameterized by a learnable quaternion $\mathbf{r}$ (for rotation) and a scaling vector $\mathbf{s}$ (for scaling), respectively:
\begin{equation}
\Sigma = R S S^\top R^\top
\label{eq:cov_decomposition}
\end{equation}

Given the view transformation matrix $W$ and the Jacobian $J$ of the affine approximation of the projection transformation, a 3D Gaussian is projected onto the 2D image plane. The resulting 2D covariance matrix $\Sigma'$ is computed as follows~\cite{zwicker2001ewa}:
\begin{equation}
\Sigma' = J W \Sigma W^\top J^\top
\label{eq:cov_projection}
\end{equation}

In addition to geometric parameters ($\mu, \mathbf{r}, \mathbf{s}$), each 3D Gaussian is augmented with two appearance properties: a learned opacity $\alpha \in [0, 1]$ and coefficients of Spherical Harmonics (SH) to represent its view-dependent color $\mathbf{c}$. The final color $\mathbf{C}$ for a given pixel is then rendered by alpha-blending the projected Gaussians, sorted by depth:
\begin{equation}
\mathbf{C} = \sum_{i \in N} \mathbf{c}_i \alpha_i \prod_{j=1}^{i-1} (1 - \alpha_j)
\label{eq:alpha_blending}
\end{equation}
where $\alpha_i$ is the final computed opacity. It is calculated by multiplying the learned opacity of the Gaussian with its 2D Gaussian function evaluated at the pixel center using the projected covariance $\Sigma'$.

\subsection{Surface-Embedded Gaussian Initialization}
Previous head representations based on implicit functions typically operate on a "canonical-plus-deformation" principle, driven by a 3DMM. This is commonly realized either by directly conditioning the implicit field on expression codes \cite{gafni2021dynamic} or by propagating vertex-wise deformations from the 3DMM's canonical space to its observed state \cite{athar2022rignerf,zielonka2023instant}. However, this strategy suffers from a critical limitation: it fails to sufficiently leverage the rich geometric priors inherent in the 3DMM mesh itself. Consequently, as demonstrated by Xu et al. \cite{xiang2024flashavatar}, such approaches struggle to robustly model dynamic avatars with complex expressions, as the resulting deformation field lacks the necessary structural awareness.

To overcome this limitation, we bind 3D Gaussians to the 3DMM mesh surface to ensure cohesive deformation with the geometry. Following \cite{xiang2024flashavatar}, UV sampling is used for initialization, forming a mesh-embedded Gaussian field that models both appearance and expression-driven deformation while maintaining a uniform point layout. Our key innovation is learning an additional offset field for non-surface details via a Cross-Attention module, which dynamically modulates conditioning signals instead of concatenating global expression codes. An overview is shown in Fig. 1.

As previously established, the Gaussian field can be parameterized by $\mathcal{G} = \{\boldsymbol{\mu}, \mathbf{r}, \mathbf{s}, \mathbf{o}, \mathbf{h}\}$.
Through UV sampling, we define the initial position of each mesh-attached Gaussian, denoted as $\boldsymbol{\mu}_M$.
In our setup, the opacity $\alpha$, SH coefficients $\mathbf{h}$, rotation $\mathbf{r}$, and scaling $\mathbf{s}$ are all learnable parameters.
While the first two attributes, which determine the primary appearance of the avatar, are optimized to converge to static values, the geometric parameters, rotation $\mathbf{r}$ and scaling $\mathbf{s}$ are learned along with an additional positional offset to $\boldsymbol{\mu}_M$.
This allows the model to capture non-surface features and fine-grained dynamic details of the face.

\begin{figure*}[!ht]
\centering
\includegraphics[width=0.6\linewidth]{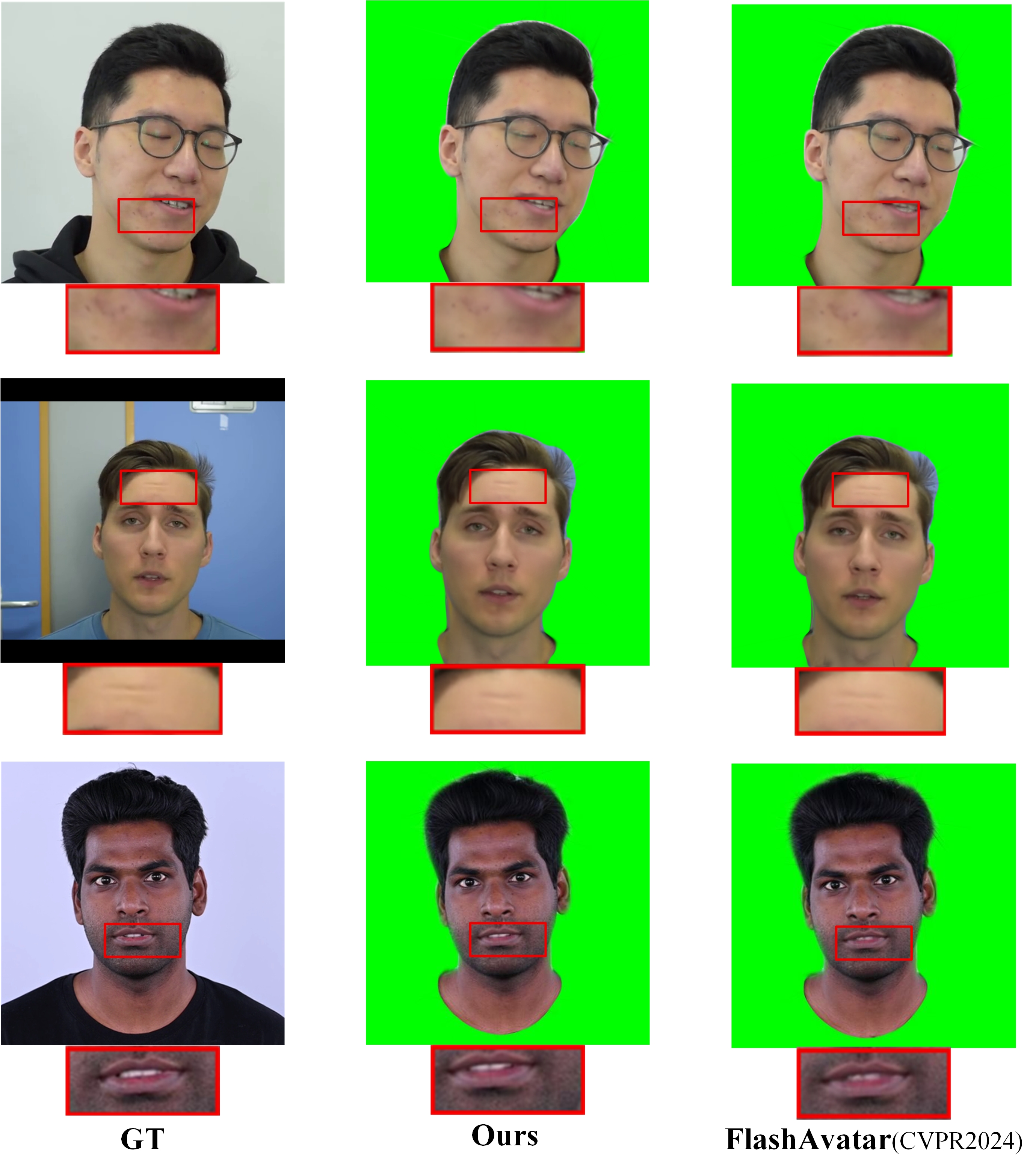}
\caption{Qualitative comparison with state-of-the-art head portrait reconstruction methods. Our model demonstrates high fidelity in reconstructing fine facial details and subtle expressions. Notably, through our proposed cross-attention fusion mechanism, our method achieves significantly improved reconstruction of challenging non-skin regions, such as teeth, compared to prior works.}
\end{figure*}

\subsection{Conditionally Adaptive Gaussian Offsets}
While attaching Gaussian primitives to a parametric facial model captures the coarse geometric motion induced by expression changes (i.e., the initial deformation from the canonical position $\boldsymbol{\mu}_T$ to the driven position $\boldsymbol{\mu}_M$), this is insufficient for representing non-surface regions (e.g., hair) and fine-grained facial dynamics (e.g., wrinkles and muscle actuations). More critically, this coarse deformation fails to distinguish between the distinct motion patterns of different components, such as the deformable skin and the rigid teeth.

To address this limitation, we move beyond simple feature concatenation strategies and design a Conditionally Adaptive Fusion Module.
The core of this module is the Cross-Attention Mechanism, which predicts a dynamic spatial offset for each Gaussian primitive.

Specifically, we treat the position of each Gaussian primitive in the canonical space, $\boldsymbol{\mu}_T$ (after being processed by a positional encoding function $\gamma$), as the \textit{Query}. The global expression code $\boldsymbol{\psi}$ serves as both the \textit{Key} and the \textit{Value}.
Through attention computation, the model generates a context-aware expression feature $\mathbf{c}_{\psi}$ for each location, which encapsulates the most relevant driving information for that specific primitive. This process is formulated as follows:
\begin{equation}
\mathbf{c}_{\psi} = \text{Attention}(\gamma(\boldsymbol{\mu}_T)W_Q, \boldsymbol{\psi}W_K, \boldsymbol{\psi}W_V)
\label{eq:cross_attention}
\end{equation}
where $\gamma$ is the positional encoding of Mildenhall et al.~\cite{mildenhall2021nerf}, and $W_Q$, $W_K$, $W_V$ are learnable projection matrices. This newly generated context feature $\mathbf{c}_{\psi}$ is then concatenated with the original positional encoding $\gamma(\boldsymbol{\mu}_T)$ and fed into a lightweight Multi-Layer Perceptron (MLP), $F_\theta$, to predict the final offset residuals for position, rotation, and scale:
\begin{equation}
(\Delta \boldsymbol{\mu}_{\psi}, \Delta \mathbf{r}_{\psi}, \Delta \mathbf{s}_{\psi}) = F_\theta(\text{concat}(\mathbf{c}_{\psi}, \gamma(\boldsymbol{\mu}_T)))
\label{eq:mlp_offset}
\end{equation}

Ultimately, the final spatial parameters of a Gaussian primitive under the target expression are determined by the sum of the coarse mesh-driven deformation and the predicted fine-grained offsets:
\begin{align}
\mu_{\psi}, \mathbf{r}_{\psi}, \mathbf{s}_{\psi} = (\mu_M \oplus \Delta\mu_{\psi}, \mathbf{r} \oplus \Delta\mathbf{r}_{\psi}, \mathbf{s} \oplus \Delta\mathbf{s}_{\psi})
\end{align}

This approach replaces the conventional "one-size-fits-all" conditioning strategy. For instance, for Gaussian primitives attached to the teeth region, our model learns to attend more to the components of the expression code $\boldsymbol{\psi}$ related to jaw articulation, while largely ignoring skin deformation signals, thus preserving their rigid motion characteristics. Conversely, for skin regions, the model can precisely extract the non-rigid deformation signals that induce wrinkles. This "tailor-made" driving signal for each primitive drastically improves the modeling of mixed dynamics and fine-grained details.


\subsection{Training Scheme and Implementation Details}
To optimize our proposed framework, we follow the composite loss function scheme adopted by ~\cite{xiang2024flashavatar}, in order to ensure both rendering fidelity and training stability. Our total loss is composed of a photometric loss $\mathcal{L}_{c}$ and a perceptual loss $\mathcal{L}_{lpips}$.

\noindent\textbf{Loss Function.} Given an expression code $\psi$, our model generates a complete set of 3D Gaussian parameters $\{ \mu_{\psi}, \mathbf{r}_{\psi}, \mathbf{s}_{\psi}, o, \mathbf{h} \}$, from which an image $\hat{I}$ is rendered.

\begin{figure*}[!h]
\centering
\includegraphics[width=0.6\linewidth]{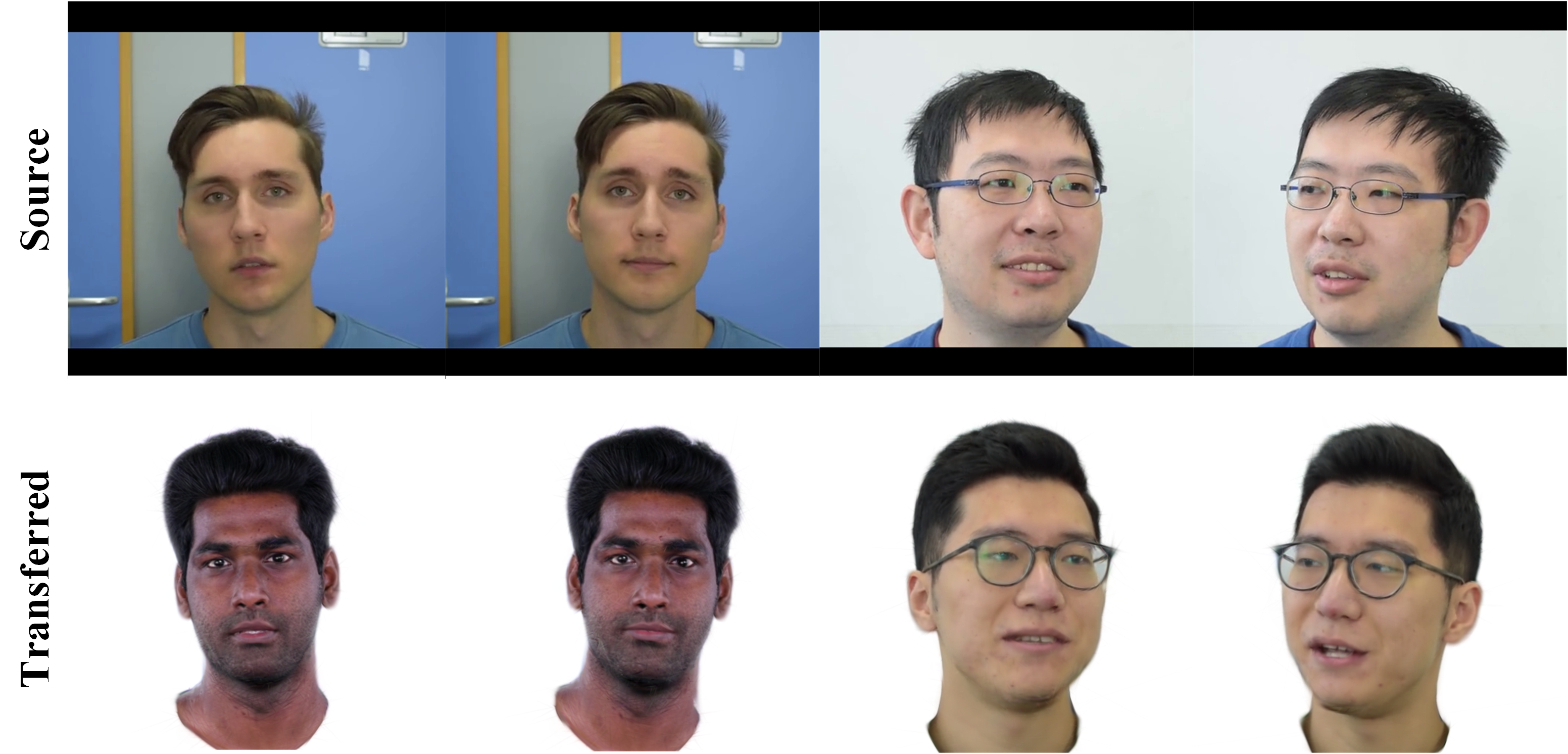}
\caption{Qualitative results of our method on the face reenactment task. Our method preserves personalized facial details in the hair, eye, and mouth regions and synthesizes more natural results.}
\end{figure*}  

To measure the pixel-level discrepancy between the rendered image $\hat{I}$ and the ground-truth image $I$, we adopt the Huber loss $\mathcal{L}_H$ (with $\delta=0.1$)~\cite{huber1992robust}, which is robust to outliers. Recognizing the challenge of reconstructing the mouth region, particularly the teeth, we introduce a mouth mask $M$ and apply a higher weight $\lambda_{mouth}$ to this area to encourage the model to capture finer oral details. The photometric loss $\mathcal{L}_{c}$ is thus defined as:
\begin{equation}
\label{eq:loss_photometric}
\mathcal{L}_{c} = \mathcal{L}_{H}(I, \hat{I}) + \lambda_{mouth}\mathcal{L}_{H}(I \cdot M, \hat{I} \cdot M)
\end{equation}

\noindent Relying solely on pixel-wise losses can lead to blurry results. To enhance visual realism, we further incorporate the LPIPS perceptual loss~\cite{zhang2018unreasonable}, $\mathcal{L}_{lpips}$. This loss measures the similarity between the rendered and ground-truth images in a deep feature space, utilizing a pre-trained VGG network~\cite{simonyan2014very}. It is instrumental in preserving high-frequency details and stabilizing the training process.

The final objective function $\mathcal{L}$ is a weighted sum of these two components:
\begin{equation}
\label{eq:loss_total}
\mathcal{L} = \mathcal{L}_{c} + \lambda_{lpips}\mathcal{L}_{lpips}
\end{equation}

\noindent\textbf{Implementation Details.} 
Our framework is implemented using PyTorch~\cite{paszke2019pytorch}.We leverage the efficient differentiable Gaussian rasterizer proposed by 3D-GS~\cite{kerbl2023gaussian} and utilize PyTorch3D~\cite{ravi2020accelerating} for the initial mesh-to-UV space mapping. For FLAME model tracking, we employ the analysis-by-synthesis face tracker from MICA~\cite{zielonka2022towards}, with further improvements inspired by INSTA~\cite{zielonka2023instant}. The expression code $\psi$ is a concatenation of the tracked expression coefficients, eye poses, jaw pose, and eyelid parameters.

\begin{figure}[h]
\centering
\includegraphics[width=0.8\linewidth]{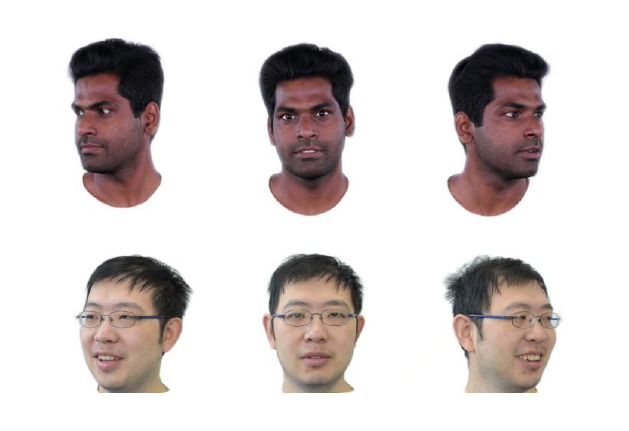}
\caption{Demonstration of new 3D consistent head pose synthesis and expression variations.}
\end{figure}

\noindent\textbf{Optimization Settings.} 
The parameters to be optimized include the color (Spherical Harmonics), opacity, initial rotation, and scale of the 3D Gaussians, as well as all trainable parameters of our proposed Conditionally Adaptive Fusion Module and the Offset MLP. We use the Adam optimizer~\cite{kingma2014adam} with $\beta=(0.9, 0.999)$. The learning rates for the Gaussian attributes are consistent with the official 3D-GS implementation, while the learning rate for our newly introduced fusion module and offset MLP is set to $\eta=1\mathrm{e}{-4}$. For our training strategy, we set $\lambda_{mouth}=40$. To stabilize the initial phase of training, the weight for the perceptual loss, $\lambda_{lpips}$, is set to 0 for the first 15,000 iterations and subsequently increased to 0.05.

\begin{table}[htbp]
\caption{Quantitative comparison with state-of-the-art head reconstruction methods}
\begin{center}
\begin{tabular}{|l|c|c|c|c|}
\hline
\textbf{Method} & \textbf{L1} $\downarrow$ & \textbf{PSNR} $\uparrow$ & \textbf{SSIM} $\uparrow$ & \textbf{LPIPS} $\downarrow$ \\
\hline
FlashAvatar (CVPR2024) & 0.0137 & 26.51 & 0.9256 & 0.0971 \\
Ours & \textbf{0.0128} & \textbf{26.53} & \textbf{0.9329} & \textbf{0.0810} \\
\hline
\end{tabular}
\label{tab:ablation}
\end{center}
\end{table}

\section{EXPERIMENTAL EVALUATION}
\subsection{Experimental Setup}
To ensure reproducibility and fair comparison, we conducted our experiments on publicly available video datasets previously used in seminal works~\cite{gafni2021dynamic, zielonka2023instant}. All input video sequences undergo a standardized pre-processing pipeline. First, the frames are cropped and uniformly resized at a resolution of $512 \times 512$. Subsequently, we employ Robust Video Matting (RVM)~\cite{lin2022robust} to extract the foreground subject. To facilitate accurate modeling of internal structures, we utilize a parsing network based on BiSeNet~\cite{yu2021bisenet} to precisely segment the oral region. The processed video clips typically range from one to three minutes in length. Following standard evaluation protocols, we reserve the final 500 frames of each sequence for the test set, with the remainder used for training. For a fair comparison, all experiments were conducted on a single NVIDIA RTX 4090 GPU. The number of training iterations for all methods is fixed at 15,000.

\subsection{Comparative Analysis}
We benchmark our method against the state-of-the-art dynamic head avatar framework, FlashAvatar~\cite{xiang2024flashavatar}, which represents the prior art of global conditioning strategy. As illustrated in Fig. 2, our method demonstrates clear superiority in reconstruction quality. While FlashAvatar~\cite{xiang2024flashavatar} suffers from blurring and geometric distortion in the dental region, our approach reconstructs sharp and structurally-accurate teeth. Furthermore, our reconstructions of high-frequency details, such as acne marks and wrinkles, exhibit higher fidelity to the ground truth. The quantitative results in Table 1 corroborate these qualitative findings. Our method significantly outperforms the baseline in all standard metrics, including PSNR, SSIM, and LPIPS, providing strong quantitative evidence for the superior fidelity of our framework. A key advantage of our framework is its precise, decoupled control over expression and pose. As shown in Fig. 3, for expression reanimation, our model accurately replicates dynamics while preserving key identity features, such as hairstyle and eyes, yielding natural results. Furthermore, Fig. 4 demonstrates the model's generalization to novel expression and pose synthesis.

\section{Conclusion}
In this paper, we addressed a key limitation in 3D Gaussian Splatting-based avatars: inaccurate local deformations from global conditioning, which causes artifacts on rigid structures like teeth. Our solution was a Conditionally Adaptive Fusion Module that used cross-attention to generate a unique, spatially-aware driving signal for each Gaussian primitive. This design effectively disentangled expression dynamics. Our framework integrated this module into 3D-GS and achieved better results than the baseline. Experiments confirmed the remarkable improvement in reconstruction fidelity, especially for teeth, while preserving real-time rendering. Our research advances the development of high-fidelity, drivable digital humans, with future work focused on full-body capture and optimized attention mechanisms for greater efficiency.

\section*{Acknowledgements}

This work was funded by UKRI (EP/Z000025/1) and the Horizon Europe Programme under the MSCA grant for ACMod (Grant No. 101130271).

\end{document}